\documentclass[12pt]{article}

\newcommand{\be}{\begin{equation}}
\newcommand{\ee}{\end{equation}}
\newcommand{\bea}{\begin{eqnarray}}
\newcommand{\eea}{\end{eqnarray}}
\newcommand{\beas}{\begin{eqnarray*}}
\newcommand{\eeas}{\end{eqnarray*}}

\begin{document}
\begin{titlepage}

\begin{center}

{\Large Local bulk physics from intersecting modular Hamiltonians}

\vspace{8mm}

\renewcommand\thefootnote{\mbox{$\fnsymbol{footnote}$}}
Daniel Kabat${}^{1}$\footnote{daniel.kabat@lehman.cuny.edu},
Gilad Lifschytz${}^{2}$\footnote{giladl@research.haifa.ac.il}

\vspace{4mm}

${}^1${\small \sl Department of Physics and Astronomy} \\
{\small \sl Lehman College, City University of New York, Bronx NY 10468, USA}

\vspace{2mm}

${}^2${\small \sl Department of Mathematics and Haifa Research Center for Theoretical Physics and Astrophysics} \\
{\small \sl University of Haifa, Haifa 31905, Israel}

\end{center}

\vspace{8mm}

\noindent
We show that bulk quantities localized on a minimal surface homologous to a boundary region correspond in the CFT  to operators that commute with the modular Hamiltonian associated with the boundary region.  If two such minimal surfaces intersect at a point in the bulk then CFT operators which commute with both extended modular Hamiltonians must be localized at the intersection point.  We use this to construct local bulk operators purely from CFT considerations, without knowing the bulk metric, using intersecting modular Hamiltonians.  For conformal field theories at zero and finite temperature the appropriate modular
Hamiltonians are known explicitly and we recover known expressions for local bulk observables.

\end{titlepage}
\setcounter{footnote}{0}
\renewcommand\thefootnote{\mbox{\arabic{footnote}}}

%%%%%%%%%%%%%%%%%%%%%%%
\section{Introduction}

Having a procedure for recovering bulk physics from the CFT is fundamental to our quest to understand quantum gravity using AdS/CFT \cite{Maldacena:1997re}. 
The Ryu-Takayanagi (RT) relation \cite{Ryu:2006bv} between the area of minimal surfaces anchored on the boundary
(RT surfaces) and entanglement entropy in the CFT is one example of how bulk quantities can be computed from the CFT. The first law of entanglement entropy has been used together with the RT and HRT formulas \cite{Hubeny:2007xt} to derive
the linearized Einstein equations in the bulk \cite{Faulkner:2013ica}. 

Natural bulk objects to consider are bulk field operators.
Traditionally the program of constructing bulk field operators in the CFT starts from knowledge of the bulk metric.
This information is used to compute smearing functions which provide the leading-order expression (in $1/N$)
for a local bulk field in terms of a single-trace primary scalar operator in the CFT \cite{Bena:1999jv, Hamilton:2005ju, Hamilton:2006az}.
\begin{equation}
\Phi^{(0)}(x,z)=\int dx' K(x,z|x'){\cal O}(x') 
\label{one}
\end{equation}
$\Phi^{(0)}$ is a CFT operator which reproduces the correct bulk 2-point function when inserted in a CFT correlator.
However the expression for $\Phi^{(0)}$ is not unique.
Among the expressions we will be using are the
complex coordinate representation (for Poincar\'e coordinates in AdS${}_3$) \cite{Hamilton:2006fh}
\be
\Phi^{(0)}(Z,X,T)=\frac{\Delta-1}{\pi} \int_{y'^2+t'^2<Z^2}dt'dy' \left (\frac{Z^2-y'^2-t'^2}{Z}\right )^{\Delta-2}{\cal O}(T+t',X+iy')
\ee
and the representation in terms of mode functions \cite{Banks:1998dd}.
\begin{equation}
\Phi^{(0)}(Z,X,T)={2^\Delta \Gamma(\Delta) \over 8 \pi^2} \int_{\vert \omega \vert > \vert k \vert} d\omega dk \, e^{-i\omega T} e^{ikX} Z(\omega^2-k^2)^{-\nu/2}J_{\nu}\big(Z\sqrt{\omega^2-k^2}\big){\cal O}(\omega,k)
\label{pomom}
\end{equation}
$1/N$ corrections to these expressions -- for example to define CFT operators that will reproduce the expected bulk 3-point functions -- can be obtained purely from the CFT using bulk locality as a guiding principle.  Imposing bulk microcausality corresponds to canceling unwanted singularities in correlation functions.  To achieve this it is necessary
to correct the definition of the bulk field by adding to $\Phi^{(0)}$ an infinite tower of higher-dimension multi-trace operators.  To restore locality in $n$-point functions it is necessary to add CFT operators involving up to $n-1$ traces, all smeared with the appropriate smearing functions as in (\ref{one}). In this way the expressions relevant to bulk 3-point functions were obtained in \cite{Kabat:2011rz, Kabat:2015swa}, and the resulting bulk fields were shown to obey the correct bulk equations of motion. The expressions needed to reproduce bulk 4-point functions were obtained in \cite{Kabat:2016zzr} with the help of crossing symmetry. These results extend with some modifications to bulk fields with spin \cite{Heemskerk:2012np, Kabat:2012hp, Kabat:2013wga, Sarkar:2014dma}. However throughout this program the starting
point, that is the lowest-order smearing functions, were computed using knowledge of the bulk metric.\footnote{An alternative construction for empty AdS is
based on representation theory \cite{Dobrev:1998md,Aizawa:2014yqa}.}

In this paper we obtain the zeroth-order bulk operator $\Phi^{(0)}$, up to a multiplicative coefficient,\footnote{The requirement of locality
only fixes bulk observables up to a multiplicative coefficient.  The coefficient could be chosen to depend on bulk space-time position.
We will discuss this ambiguity more in section 3.} purely from CFT considerations.
The basic idea is that the modular Hamiltonian associated to a boundary region has the same action on bulk quantities as
the associated bulk modular Hamiltonian \cite{Jafferis:2015del}.  The RT surface (the minimal bulk surface homologous
to the boundary region) plays the role of a bifurcation surface in Rindler coordinates (the surface where the past and future Rindler horizons
intersect).  Just as the bifurcation surface is invariant under Rindler time evolution, the RT surface is invariant under the action of the bulk modular Hamiltonian.
This means CFT objects which are localized in the bulk on an RT surface should commute with the corresponding boundary modular Hamiltonian.

To proceed it's convenient to define an extended modular Hamiltonian which generates a non-trivial flow everywhere
in the bulk
except on the RT surface.  If a collection of RT surfaces intersect at a point in the bulk, then CFT quantities which commute with all of the
corresponding extended modular Hamiltonians must be localized at the intersection point.  We can impose this as a condition to
construct local observables in the bulk. The construction is simplest in the case of 2-D CFT where RT surfaces are just lines. 

Recently $\Phi^{(0)}$ has been shown to be related to several natural CFT objects. In \cite{Czech:2016xec, daCunha:2016crm, deBoer:2016pqk} the integral of $\Phi^{(0)}$ over an RT surface was found to be related to a conformal block operator.\footnote{There is also a connection with geodesic Witten diagrams \cite{Hijano:2015zsa}.} In \cite{Miyaji:2015fia, Verlinde:2015qfa, Nakayama:2015mva} it was shown that $\Phi^{(0)}$ creates a boundary cross-cap state in the CFT. One of the motivations of the present paper was to understand the relation between these different descriptions of $\Phi^{(0)}$. The construction developed here makes the connection much clearer.

An outline of this paper is as follows.  In section 2 we review the modular Hamiltonian appropriate to a segment of the boundary and extend it outside this region in a natural way to obtain what we call the extended modular Hamiltonian. We use this to show that bulk operators on the RT surface commute with the modular Hamiltonian in a variety of contexts. In section 3 we turn the argument around and search for a CFT operator that commutes with the extended modular Hamiltonians associated with two different boundary segments. Provided the corresponding RT surfaces intersect we explicitly solve this condition and find the correct smearing function for a local operator that lives on the intersection.
We do this for the vacuum state of the CFT on a line in three ways: using complexified coordinates, using a derivative expansion, and in momentum space.
The latter recovers the Poincar\'e mode expansion of a bulk field from the CFT. We also do this for a CFT on a line at finite temperature and recover the complexified smearing function for a BTZ black hole.  In appendix A we establish the relationship
between conventional and extended modular Hamiltonians, in appendix B we show that the extended modular Hamiltonian has the same action on bulk operators which are off the RT surface as the bulk Rindler Hamiltonian, and in appendix C we study geodesics in BTZ.  For convenience in all explicit computations we specialize to AdS${}_3$/CFT${}_2$.

%%%%%%%%%%%%%%%%%%%%%%%%%%%%%%%%%
\section{Modular Hamiltonian in AdS${}_3$}

We work in AdS${}_3$ / CFT${}_2$ in Poincar\'e coordinates. The bulk metric is
\begin{equation}
ds^2=\frac{l^2}{Z^2}(-dT^2+dZ^2+dX^2)
\end{equation}
and the CFT metric is
\begin{equation}
ds^2=-dT^2+dX^2
\end{equation}
We introduce boundary light-front coordinates
\begin{equation}
\xi=X-T, \ \  \bar{\xi}=X+T
\end{equation}
A space-like segment in a $(1+1)$ dimensional CFT defines a causal diamond based on the segment. The diamond $D(x,y)$ is defined through its upper tip $y^{\mu}$ and its lower tip $x^{\mu}$, which we describe using the light-front coordinates
\begin{equation}
\label{uv}
(u,\bar{u})=(x^1-x^0,x^1+x^0), \ \ (v,\bar{v})=(y^1-y^0, y^1+y^0)
\end{equation}
If we choose a diamond on the boundary whose left and right tips lie on the $T=0$ slice at points $y_1$ and $y_2$ then
\be
y^{\mu}=\left(\frac{y_2 - y_1}{2}, \frac{y_2 + y_1}{2}\right) \qquad x^{\mu}=\left(\frac{y_1 - y_2}{2}, \frac{y_2 + y_1}{2}\right)
\ee
so that
\begin{equation}
u=y_2, \ \ v=y_1, \ \  \bar{u}=y_1, \ \ \bar{v}=y_2\,.
\end{equation}
For a CFT in its vacuum state the modular Hamiltonian can be written explicitly \cite{Casini:2011kv}.
\begin{equation}
H_{mod}=2\pi\int_{v}^{u}d\xi \frac{(u-\xi)(\xi-v)}{u-v}T_{\xi\xi}(\xi)+2\pi \int_{\bar{u}}^{\bar{v}}d\bar{\xi} \frac{(\bar{v}-\bar{\xi})(\bar{\xi}-\bar{u})}{\bar{v}-\bar{u}}\bar{T}_{\bar{\xi}\bar{\xi}}(\bar{\xi})
\label{hmod}
\end{equation}
We define coordinates $\eta$ and $\bar{\eta}$ by
\begin{eqnarray}
d\eta & = & \frac{(u-v)}{(u-\xi)(\xi-v)} d\xi \nonumber \\
d\bar{\eta} & = & \frac{(\bar{v}-\bar{u})}{(\bar{v}-\bar{\xi})(\bar{\xi}-\bar{u})} d\bar{\xi}
\label{wxi}
\end{eqnarray}
which are solved by
\begin{eqnarray}
e^{\eta}&=&\frac{\xi-v}{u-\xi} \nonumber\\
e^{\bar{\eta}}&=&\frac{\bar{\xi}-\bar{u}}{\bar{v}-\bar{\xi}}
\label{rineta}
\end{eqnarray}
These Rindler-like null coordinates $(\eta,\bar{\eta})$ cover the diamond. In terms of dimensionless time and space coordinates
\begin{equation}
\eta=\phi-\hat{t},\ \ \bar{\eta}=\phi+\hat{t}
\end{equation}
the diamond is covered by $-\infty < \hat{t},\,\phi < \infty$.

Under the change of coordinates $\xi \rightarrow \eta$ and $\bar{\xi} \rightarrow \bar{\eta}$ we have\footnote{This is
more commonly written in terms of $T(\xi)=-2\pi T_{\xi \xi}$ and $\bar{T}(\bar{\xi})=-2\pi T_{\bar{\xi}\bar{\xi}}$.}
\begin{eqnarray}
T_{\eta\eta}(\eta)=\Big(\frac{d\xi}{d \eta}\Big)^2 T_{\xi \xi}-\frac{c}{24\pi}S(\xi,\eta) \nonumber\\
T_{\bar{\eta}\bar{\eta}}(\bar{\eta})=\Big(\frac{d\bar{\xi}}{d \bar{\eta}}\Big)^2 T_{\bar{\xi} \bar{\xi}}-\frac{\bar{c}}{24\pi}S(\bar{\xi},\bar{\eta})
\end{eqnarray}
where
\begin{equation}
S(\xi, \eta)=\frac{\frac{d\xi}{d \eta}\frac{d^3 \xi}{d\eta^3}-\frac{3}{2}\Big(\frac{d^2\xi}{d\eta^2}\Big)^2}{\Big(\frac{d\xi}{d \eta}\Big)^2}
\end{equation}
For the change of coordinates in (\ref{wxi}) one finds $S(\xi,\eta)=S(\bar{\xi}, \bar{\eta})=-\frac{1}{2}$.
Thus we can write the modular Hamiltonian (\ref{hmod}) as
\begin{equation}
H_{mod}=2\pi \int_{-\infty}^{\infty} d\phi (T_{\eta\eta}(\phi) + T_{\bar{\eta} \bar{\eta}}(\phi))-\frac{c}{24}\int_{y_1}^{y_2} d \xi \frac{y_2-y_1}{(y_2-\xi)(\xi-y_1)}-\frac{\bar{c}}{24}\int_{y_1}^{y_2} d \bar{\xi} \frac{y_2-y_1}{(y_2-\bar{\xi})(\bar{\xi}-y_1)}
\label{hmodrt}
\end{equation}
In Poincar\'e coordinates a bulk geodesic $\gamma$ on the $T = 0$ slice connecting the points $y_1$ and $y_2$ on the boundary is given by the semicircle
\begin{equation}
(X-y_1)(y_2-X)=Z^2
\label{poingeo}
\end{equation}
Proper length along this geodesic is
\begin{equation}
ds=\frac{l}{2}\frac{y_2-y_1}{(X-y_1)(y_2-X)}dX
\label{ds}
\end{equation}
Using this and $c=\bar{c}=\frac{3l}{2G}$ the last two terms in (\ref{hmodrt}) can be seen to be
\begin{equation}
-\frac{1}{4G} \int_{\gamma} ds
\end{equation}
This is just the RT term, i.e.\ the area of the minimal surface. So in fact
\begin{equation}
2\pi \int_{-\infty}^{\infty} d\phi (T_{\eta\eta}(\phi) + T_{\bar{\eta} \bar{\eta}}(\phi))=H_{mod}+\frac{A}{4G}
\label{jlms}
\end{equation}
The authors of \cite{Jafferis:2015del} identified the left-hand side of (\ref{jlms}) with the boundary modular Hamiltonian and interpreted $H_{mod}$ as a bulk modular Hamiltonian which generates bulk time evolution in the appropriate bulk Rindler wedge plus fluctuations of the RT surface (see also \cite{Czech:2016tqr}).
The computations of this paper will, among other things,  confirm that $H_{mod}$ acts on CFT operators which represent bulk quantities in the manner expected for a bulk Rindler Hamiltonian.

%%%%%%%%%%%%%%%%%%%%%%%%%%%%%%%%%%%%%%%%%%%%%%%
\subsection{CFT quantities invariant under the modular Hamiltonian}

We start with the expression for a local bulk operator in Rindler coordinates \cite{Hamilton:2006az},
\begin{equation}
\label{RindlerSmear}
\Phi^{(0)}(r,\phi,t)=\frac{(\Delta-1)2^{\Delta-2}}{\pi r_{+}^{\Delta}}\int dxdy\left ( \frac{r}{r_{+}} \left (\cos y-\sqrt{1-\frac{r_{+}^{2}}{r^2}}\cosh x \right ) \right )^{\Delta-2}{\cal O}_{Rindler}(\phi+i\frac{ly}{r_{+}}, t+\frac{l^2x}{r_+})
\end{equation}
where the AdS${}_3$ metric is
\begin{equation}
ds^2=-\frac{r^2-r^{2}_{+}}{l^2}dt^2+\frac{l^2}{r^2-r_{+}^2}dr^2 +r^2 d\phi^2
\end{equation}
Here $l$ is the AdS radius, $r_{+}$ is the horizon radius and the region of integration is
\begin{equation}
\cos y >\sqrt{1-\frac{r_{+}^{2}}{r^2}}\cosh x.
\end{equation}
The Rindler operator in the CFT is normalized according to $\lim_{r \rightarrow \infty} r^\Delta \Phi^{(0)}(r,\phi,t) = {\cal O}_{Rindler}(\phi,t)$.
We understand the analytic continuation to complex boundary coordinates to be defined by
\begin{equation}
{\cal O}_{Rindler}(\phi+i\frac{ly}{r_{+}}, t+\frac{l^2x}{r_+})=\int d\omega dk \, e^{-i\omega (t+\frac{l^2x}{r_{+}})}e^{ik(\phi+i\frac{ly}{r_{+}})}{\cal O}_{Rindler}(\omega,k)
\end{equation}
As $r \rightarrow r_{+}$ the integration region becomes $-\infty <x< \infty$  and $-\frac{\pi}{2}<y<\frac{\pi}{2}$. Thus
\begin{equation}
\Phi^{(0)}(r_{+},\phi,t)=\frac{(\Delta-1)2^{\Delta-2}}{\pi r_{+}^{\Delta}}\int_{-\frac{\pi}{2}}^{\frac{\pi}{2}}dy \cos^{\Delta-2} y \int d\omega dk e^{-i\omega t}e^{ik(\phi+i\frac{ly}{r_{+}})}{\cal O}_{Rindler}(\omega,k)\int_{-\infty}^{\infty} dx e^{-i\omega \frac{l^2 x}{r_{+}}}
\end{equation}
The integral over $x$ sets $\omega$ to zero.  Then using
\begin{equation}
\int_{-\frac{\pi}{2}}^{\frac{\pi}{2}} dy \cos^{\Delta-2} y \  e^{-kly/r_{+}} = \frac{\Gamma(\Delta)}{2^{\Delta-1} (\Delta-1)}\frac{1}{|\Gamma(\frac{\Delta}{2}+\frac{ikl}{2r_{+}})|^2}
\end{equation}
and defining
\begin{equation}
\Phi^{(0)}(r_{+},k,t)=\frac{1}{2\pi}\int_{-\infty}^{\infty}d\phi e^{-ik\phi} \Phi^{(0)}(r,\phi,t)
\end{equation}
we get
\begin{equation}
\Phi^{(0)}(r_{+},k,t)=\frac{\Gamma(\Delta)}{l r_{+}^{\Delta-1}}\frac{1}{|\Gamma(\frac{\Delta}{2}+\frac{ikl}{2r_{+}})|^2}{\cal O}_{Rindler}(\omega=0,k)
\label{modmod}
\end{equation}

This shows that zero frequency modes relative to the boundary Rindler Hamiltonian live on the bulk RT surface, and due to (\ref{jlms}), that bulk objects on the RT surface commute with  $H_{mod}$.
Setting $k = 0$ in (\ref{modmod}) gives
\begin{equation}
{1 \over 2 \pi} \int_{-\infty}^{\infty} d\phi \, \Phi^{(0)}(r,\phi,t)=\frac{\Gamma(\Delta)}{ l r_{+}^{\Delta-1} \Gamma^{2}(\frac{\Delta}{2})}{\cal O}_{Rindler}(\omega=0,k=0)
\label{zerod2bulk}
\end{equation}
where the left-hand side is up to a constant the integral of the bulk field over the RT surface which serves as the
horizon of the bulk Rindler wedge.

In fact (\ref{zerod2bulk}) follows from results in the literature.
The integral of a bulk field operator over an RT surface $\gamma$ was identified in \cite{Czech:2016xec, deBoer:2016pqk} with a particular CFT expression.
In two dimensions, for a primary operator with dimensions $h=\bar{h}=\frac{1}{2}\Delta_{\cal O}$, the appropriate identification was found to be
\begin{eqnarray}
&& Q({\cal O};u,\bar{u;v,\bar{v}}) = {C_{\rm blk} \over 8 \pi G_N} \int_\gamma ds \, \Phi^{(0)} \nonumber \\
&& \qquad = C_{\cal O}\int_{D(x,y)} d\xi d\bar{\xi} \left (\frac{(u-\xi)(\xi-v)}{(u-v)}\right )^{h-1}
\left (\frac{(\bar{v}-\bar{\xi})(\bar{\xi}-\bar{u})}{(\bar{v}-\bar{u})}\right )^{\bar{h}-1}{\cal O}(\xi,\bar{\xi})
\label{zerod2} 
\end{eqnarray}
where $C_{\rm blk}$ and $C_{\cal O}$ are normalization constants.
To see what the right hand side of (\ref{zerod2}) represents, we make a conformal transformation $\xi \rightarrow \eta$ and $\bar{\xi} \rightarrow \bar{\eta}$ as in (\ref{rineta})
and define a Rindler operator ${\cal O}_R$ as the conformal transformation\footnote{The relation between ${\cal O}_{Rindler}$ and ${\cal O}$ is usually taken to be
${\cal O}_{Rindler}=\lim_{r \rightarrow \infty}(rZ)^{\Delta}{\cal O}$. See (39) in \cite{Hamilton:2006fh}.  This normalization gives an extra factor of $r_{+}^{\Delta}$ compared to (\ref{normrp}),
so that ${\cal O}_{Rindler} = r_{+}^{\Delta}{\cal O}_{R}$.} of ${\cal O}$
\begin{equation}
{\cal O}_{R}(\phi,\hat{t})=\left (\frac{(u-\xi)(\xi-v)}{(u-v)}\right )^{h}
\left (\frac{(\bar{v}-\bar{\xi})(\bar{\xi}-\bar{u})}{(\bar{v}-\bar{u})}\right )^{\bar{h}}{\cal O}(\xi,\bar{\xi})
\label{normrp}
\end{equation}
Then we see that 
\begin{eqnarray}
Q({\cal O};u,\bar{u};v,\bar{v}) & = & C_{\cal O}\int_{-\infty}^{\infty} \int_{-\infty}^{\infty} d\hat{t} d\phi \, {\cal O}_{R}(\hat{t},\phi) \nonumber \\
& = & C_{\cal O}{\cal O}_{R}(\omega=0,k=0)
\label{zerod2f}
\end{eqnarray}
which up to constants agrees with (\ref{zerod2bulk}).

It will be useful below to show directly that $Q$ commutes with $H_{mod}$.
In Lorentzian CFT the commutator of the energy-momentum tensor with a primary field is
\begin{eqnarray*}
2\pi[T_{ww}(w),{\cal O}] &=& 2\pi i(h\partial_{\xi} \delta (\xi-w) {\cal O}+\delta (\xi-w) \partial_{\xi}{\cal O})\\
2\pi[ T_{\bar{w}\bar{w}}(\bar{w}),{\cal O}] &=& -2\pi i(\bar{h}\partial_{\bar{\xi}}\delta(\bar{\xi}-\bar{w}) {\cal O}+\delta(\bar{\xi}-\bar{w})\partial_{\bar{\xi}}{\cal O})
\end{eqnarray*}
Now we can compute the action of the modular Hamiltonian on a CFT operator. 
We start with the modular Hamiltonian for a segment $(y_1,y_2)$ on the $T=0$ time slice, $H_{mod} = H_{mod}^{(R)} + H_{mod}^{(L)}$ where
\begin{eqnarray}
H_{mod}^{(R)}&=&2\pi \int_{y_1}^{y_2} \frac{(w-y_1)(y_2-w)}{y_2-y_1} T_{ww}(w) \label{Hmoddef} \\
\nonumber H_{mod}^{(L)}&=&2\pi \int_{y_1}^{y_2} \frac{(\bar{w}-y_1)(y_2-\bar{w})}{y_2-y_1} T_{\bar{w}\bar{w}}(\bar{w})
\end{eqnarray}
We find
\begin{eqnarray}
&& [H_{mod}^{(R)}, {\cal O}(\xi,\bar{\xi})]=\Theta((\xi-y_1)(y_2-\xi)) \frac{2 \pi i}{y_2-y_1}\left ( h(y_2+y_1-2\xi)+(\xi-y_1)(y_2-\xi)\partial_{\xi}\right ) {\cal O}(\xi,\bar{\xi}) \nonumber\\
\label{hmodregr} \\
&& [H_{mod}^{(L)}, {\cal O}(\xi,\bar{\xi})]=-\Theta((\bar{\xi}-y_1)(y_2-\bar{\xi}))\frac{2 \pi i}{y_2-y_1}\left ( h(y_2+y_1-2\bar{\xi})+(\bar{\xi}-y_1)(y_2-\bar{\xi})\partial_{\bar{\xi}}\right ) {\cal O}(\xi,\bar{\xi}) \nonumber
\end{eqnarray}
One can then easily check that $Q({\cal O};u,\bar{u;v,\bar{v}})$ commutes with $H_{mod}$.  In fact $Q$ is the unique expression which commutes with both $H_{mod}^{(R)}$ and
$H_{mod}^{(L)}$. To see this act on (\ref{zerod2}) using (\ref{hmodregr}) and integrate by parts.

In fact there are generalizations of  (\ref{zerod2f}). A mode of the boundary Rindler operator with zero frequency but
non-zero momentum $Q_k \equiv {\cal O}_{R}(\omega=0,k)$ can be written as
\be
Q_{k}= 
\int_{D(x,y)} d\xi d\bar{\xi} \left (\frac{(\xi-v)(\bar{\xi}-\bar{u})}{(u-\xi)(\bar{v}-\bar{\xi})}\right )^{\frac{ik}{2}}\left (\frac{(u-\xi)(\xi-v)}{(u-v)}\right )^{h-1}
\left (\frac{(\bar{v}-\bar{\xi})(\bar{\xi}-\bar{u})}{(\bar{v}-\bar{u})}\right )^{\bar{h}-1}{\cal O}(\xi,\bar{\xi})
\ee
This commutes with the sum $H_{mod}^{(R)}+H_{mod}^{(L)}$ but obeys $[H_{mod}^{(R)}-H_{mod}^{(L)}, Q_{k}]=2\pi k Q_{k}$. From (\ref{modmod}) $Q_k$ is related to the integral of $\Phi^{(0)}$ over the RT surface with a particular weight.

%%%%%%%%%%%%%%%%%%%%%%%%%%%
\subsection{Extended modular Hamiltonian}

In what follows we will want to compare the action of two modular Hamiltonians based on different segments
of the boundary. To make this comparison it is very convenient to define what we call an extended modular Hamiltonian $\tilde{H}_{mod}$.\footnote{This quantity has appeared before in the literature.
In \cite{Jafferis:2015del,Jafferis:2014lza} $\tilde{H}_{mod}$ was referred to as the total modular operator $K$, and in \cite{Faulkner:2016mzt} it was referred to as the full modular operator $\widehat{K}$.}
The extended modular Hamiltonian agrees with the usual modular Hamiltonian within its defining segment, but it extends
in a natural way to be non-zero outside the segment. Thus the action of $\tilde{H}_{mod}$ on operators inside the diamond $D(x,y)$ based
on the segment will be the same as the action of the usual modular Hamiltonian, but $H_{mod}$
and $\tilde{H}_{mod}$ act differently on operators outside the diamond.

A convenient definition of the extended modular Hamiltonian for a segment $(y_1,y_2)$ of the boundary at $T=0$ is just
\begin{eqnarray}
\nonumber \tilde{H}_{mod}^{(R)}&=&2\pi \int_{-\infty}^{\infty} \frac{(w-y_1)(y_2-w)}{y_2-y_1} T_{ww}(w)\\
\tilde{H}_{mod}^{(L)}&=&2\pi \int_{-\infty}^{\infty} \frac{(\bar{w}-y_1)(y_2-\bar{w})}{y_2-y_1} T_{\bar{w}\bar{w}}(\bar{w})
\label{Extmoddef}
\end{eqnarray}
Compared to the usual definition (\ref{Hmoddef}) all we've done is extend the limits of integration.
This definition of the extended modular Hamiltonian has a natural interpretation.  As we show in appendix A, $\tilde{H}_{mod}$ can be identified with the modular Hamiltonian for an interval $A$ on the boundary minus the modular Hamiltonian for
its complement $\bar{A}$.\footnote{We are grateful to Michal Heller for suggesting this connection.}
\begin{equation}
\tilde{H}_{mod,A}=H_{mod, A}-H_{mod,\bar{A}}
\end{equation}
This has the nice feature that $\tilde{H}_{mod,A}$ generates a non-trivial flow everywhere in the bulk, except on the RT surface associated with $A$ which it leaves invariant.
This means operators which commute with $\tilde{H}_{mod}$ must be localized on the RT surface.
It follows from the definition that
\begin{eqnarray}
&& [\tilde{H}_{mod}^{(R)}, {\cal O}(\xi,\bar{\xi})]=\frac{2 \pi i}{y_2-y_1}\left ( h(y_2+y_1-2\xi)+(\xi-y_1)(y_2-\xi)\partial_{\xi}\right ) {\cal O}(\xi,\bar{\xi}) \\
&& [\tilde{H}_{mod}^{(L)}, {\cal O}(\xi,\bar{\xi})]=-\frac{2 \pi i}{y_2-y_1}\left ( h(y_2+y_1-2\bar{\xi})+(\bar{\xi}-y_1)(y_2-\bar{\xi})\partial_{\bar{\xi}}\right ) {\cal O}(\xi,\bar{\xi})
\end{eqnarray}
The action of the extended total modular Hamiltonian $\tilde{H}_{mod}^{(L)}+\tilde{H}_{mod}^{(R)}$ on a primary field is
\begin{equation}
[\tilde{H}_{mod}, {\cal O}(\xi ,\bar{\xi})]=\frac{2\pi i}{y_2-y_1}\left ( (\bar{\xi}-\xi)\Delta -y_1y_2(\partial_{\xi}-\partial_{\bar{\xi}})+(y_1+y_2)(\xi \partial_{\xi}-\bar{\xi}\partial_{\bar{\xi}})+\bar{\xi}^{2}\partial_{\bar{\xi}}-\xi ^2\partial_{\xi}\right ) {\cal O}
\label{tothamcor}
\end{equation}
Compared to the action of the usual modular Hamiltonian (\ref{hmodregr}), the only change is that there are no step functions.

Let us now look at a local bulk operator in the Poincar\'e patch and show that it commutes with the extended modular Hamiltonian appropriate for a segment whose RT surface passes through the bulk point.
The bulk operator in Poincar\'e coordinates can be written using the complexified smearing function as 
\be
\label{comprepcor}
\Phi(Z,X,T)=\frac{\Delta-1}{\pi} \int_{y'^2+t'^2<Z^2}\left (\frac{Z^2-y'^2-t'^2}{Z}\right )^{\Delta-2}{\cal O}(T+t',X+iy')
\ee
We understand the complexified spatial coordinate as corresponding to the formal expression
\be
\Phi(Z,X,T=0)=\frac{\Delta-1}{\pi} \int_{y'^2+t'^2<Z^2}\left (\frac{Z^2-y'^2-t'^2}{Z}\right )^{\Delta-2}e^{iy'\frac{d}{dX}} {\cal O}(t',X)
\ee
Then ($\xi=X-t'$, $\bar{\xi}=X+t'$)
\begin{eqnarray}
& & [\tilde{H}_{mod},\Phi]=\frac{2i(\Delta-1)}{y_2-y_1} \int_{y'^2+t'^2<Z^2}\left (\frac{Z^2-y'^2-t'^2}{Z}\right )^{\Delta-2}e^{iy'(\frac{d}{d\xi}+\frac{d}{d\bar{\xi}})} \nonumber\\
& & \left ( (\bar{\xi}-\xi)\Delta -y_1y_2(\partial_{\xi}-\partial_{\bar{\xi}})+(y_1+y_2)(\xi \partial_{\xi}-\bar{\xi}\partial_{\bar{\xi}})+\bar{\xi}^{2}\partial_{\bar{\xi}}-\xi ^2\partial_{\xi}\right ) {\cal O}(\xi,\bar{\xi})\nonumber\\
\end{eqnarray}
We now define $q=\xi+iy'$ and $p=\bar{\xi}+iy'$, so the above expression is
\begin{eqnarray}
& & [\tilde{H}_{mod},\Phi]=\frac{2i(\Delta-1)}{y_2-y_1} \int_{Z^2+(q-X)(p-X)>0}\left (\frac{Z^2+(q-X)(p-X)}{Z}\right )^{\Delta-2} \nonumber\\
& & \left ( (p-q)\Delta -y_1y_2(\partial_{q}-\partial_{p})+(y_1+y_2)(q \partial_{q}-p\partial_{p})+p^{2}\partial_{p}- q^2\partial_{q}\right ) {\cal O}(q,p)\nonumber\\
\label{modhcomp}
\end{eqnarray}
Now we can integrate by parts and after a little algebra we find that 
\begin{equation}
[\tilde{H}_{mod},\Phi(Z,X,T=0)]=0
\end{equation}
provided that
\begin{equation}
Z^2-(y_1+y_2)X+y_1y_2+X^2=0
\end{equation}
This is simply the condition that the bulk point $(Z,X,T=0)$ lies on a spacelike geodesic whose endpoints
hit the boundary at $(T=0,y_1)$ and $(T=0,y_2)$.  See (\ref{poingeo}).

%%%%%%%%%%%%%%%%%%%%%%%%%%%%%%%%%%%%%%%%%%%%%%%
\subsection{Finite temperature}

In this section we extend the previous discussion to treat a modular Hamiltonian which is not constructed from the ground state of the CFT.  Instead we consider a CFT at finite temperature.

For a CFT on a line at finite temperature $\beta^{-1}=\frac{r_{+}}{2\pi l^2}$ the modular Hamiltonian for a region $(-R,R)$  is given by \cite{Wong:2013gua, Cardy:2016fqc}
\begin{equation}
{H}_{mod}= c\left (\int_{-R}^{R} (\cosh \frac{r_{+} R}{l^2} -\cosh\frac{r_{+} \xi}{l^2})T_{\xi \xi}(\xi)+\int_{-R}^{R} (\cosh \frac{r_{+} R}{l^2} -\cosh\frac{r_{+}\bar{\xi}}{l^2})T_{\bar{\xi} \bar{\xi}}(\bar{\xi})\right )
\label{hmodther}
\end{equation}
with $c=\frac{2l^2}{r_{+}} \sinh \frac{r_{+} R}{l^2}$.  The extended modular Hamiltonian for the same region is then given by
\begin{equation}
\tilde{H}_{mod}= c\left (\int_{-\infty}^{\infty} (\cosh \frac{r_{+} R}{l^2} -\cosh\frac{r_{+} \xi}{l^2})T_{\xi \xi}(\xi)+\int_{-\infty}^{\infty} (\cosh \frac{r_{+} R}{l^2} -\cosh\frac{r_{+}\bar{\xi}}{l^2})T_{\bar{\xi} \bar{\xi}}(\bar{\xi})\right )
\label{hmodtherext}
\end{equation}
The action of the extended Hamiltonian on a primary scalar operator of dimension $2h$ is
\begin{eqnarray}
[\tilde{H}_{mod},{\cal O}]&=&c\Big[-\frac{r_{+}h}{l^2}\sinh \frac{r_{+}\xi}{l^2}+\frac{r_{+}h}{l^2}\sinh \frac{r_{+}\bar{\xi}}{l^2} \\
&+&\big(\cosh \frac{r_{+}R}{l^2} -\cosh\frac{r_{+}\xi}{l^2}\big)\partial_{\xi}-\big(\cosh \frac{r_{+}R}{l^2} -\cosh\frac{r_{+}\bar{\xi}}{l^2}\big)\partial_{\bar{\xi}}\Big]{\cal O}(\xi,\bar{\xi}) \nonumber
\end{eqnarray}

We want to show that this modular Hamiltonian commutes with a local bulk operator on the corresponding RT surface.
The bulk operator has a representation using complexified coordinates as in (\ref{RindlerSmear}),
\begin{equation}
\Phi^{(0)}(r,\phi,t=0)\sim\left(\frac{r}{r_{+}}\right)^{\Delta-2} \int dx dy \left(\cos y-\sqrt{1-\frac{r^{2}_{+}}{r^2}}\cosh x\right)^{\Delta-2}
{\cal O}(\phi+\frac{ily}{r_{+}},\frac{l^2 x}{r_{+}})
\end{equation}
where both $x,y$ are real and the region of integration is $\cos y > \sqrt{1-\frac{r^{2}_{+}}{r^2}}\cosh x$.  We understand the
operator at complex boundary coordinates to be defined by
\begin{equation}
{\cal O}(\phi+\frac{ily}{r_{+}},\frac{l^2 x}{r_{+}})=e^{i\frac{ly}{r_{+}}\frac{d}{d \phi}}{\cal O}(\phi,\frac{l^2 x}{r_{+}}).
\label{compcor}
\end{equation}
Using (\ref{hmodtherext}) and (\ref{compcor}) we get ($q=l\phi-\frac{l^2 x}{r_{+}}+i\frac{l^2y}{r_{+}}$ and $p=l\phi+\frac{l^2 x}{r_{+}}+i\frac{l^2y}{r_{+}}$) 
\begin{eqnarray}
&&[ \tilde{H}_{mod}, \Phi^{(0)}(r,\phi,t=0) ] \\
&& \sim c \left(\frac{r}{r_{+}}\right)^{\Delta - 2} \int dx dy\left (\cosh\big(\frac{r_{+}}{l^2}\frac{p+q-2l\phi}{2}\big)-\sqrt{1-\frac{r^{2}_{+}}{r^2}}\cosh\big(\frac{r_{+}}{l^2}\frac{p-q}{2}\big)\right)^{\Delta-2} \times \nonumber\\
&&\left (\frac{r_+}{l^2}h(\sinh \frac{r_+}{l^2}p-\sinh \frac{r_+}{l^2}q)+(\cosh \frac{r_{+} R}{l^2}-\cosh \frac{r_{+} q}{l^2})\partial_{q}-(\cosh \frac{r_{+} R}{l^2}-\cosh \frac{r_{+} p}{l^2})\partial_{p}\right ){\cal O}(q,p)\nonumber
\end{eqnarray}
After integrating by parts and a little algebra one finds that  
\begin{equation}
[\tilde{H}_{mod}, \Phi^{(0)}(r,\phi,t=0)]=0
\end{equation}
provided
\begin{equation}
\sqrt{1-\frac{r_{+}^{2}}{r^2}}=\frac{\cosh\frac{r_{+}}{l}\phi}{\cosh \frac{r_{+}}{l^2}R}
\end{equation}
As shown in appendix C, this condition is satisfied provided the bulk point $(r,\phi,t=0)$ lies
on a spacelike geodesic connecting the two boundary points
$(t=0,l\phi=-R)$ and $(t=0, l\phi=R)$.

%%%%%%%%%%%%%%%%%%%%%%%%%%%%%%%%%%%%%%%%%%%%%%%%%%
\section{Bulk operators from intersecting modular Hamiltonians}

We saw that a bulk operator $\Phi$ living on the RT surface associated with a segment of the boundary commutes
with the modular Hamiltonian appropriate to that segment. Of course this does not imply that $\Phi$ is local in the bulk.  But if
there is another segment on the boundary whose RT surface intersects the RT surface of the first one at a point, then
we can demand that $\Phi$ commutes with both modular Hamiltonians.  In this case $\Phi$ must be a local bulk operator living on the
intersection point.

To make a connection to other work, note that on a formal level the action of the extended modular Hamiltonian appropriate for the vacuum state of a CFT on a CFT primary given in (\ref{tothamcor}) identifies it as 
\begin{equation}
\tilde{H}_{mod}=\frac{2\pi}{y_2-y_1}(Q_{0}+y_1 y_2P_{0}+(y_1+y_2)M_{01})
\end{equation}
Here $Q_{0}$, $P_{0}$, $M_{01}$ are generators of the conformal group.
\begin{equation}
Q_{0}=i(\bar{L}_{1}-L_{1}) \ \ \ P_{0}=i(\bar{L}_{-1}-L_{-1}),\ \ M_{01}=i(\bar{L}_{0}-L_{0})
\end{equation}
So given two segments of the boundary $(y_1,y_2)$ and $(y_3,y_4)$, the condition for a bulk operator to live on the intersection of the
corresponding RT surface becomes
\begin{equation}
[(Q_{0}+y_1 y_2P_{0}+(y_1+y_2)M_{01}), \Phi ]=0 \  \ ,\  \ [(Q_{0}+y_3 y_4 P_{0}+(y_3+y_4)M_{01}),\Phi ]=0
\label{cormodeq}
\end{equation}
In  \cite{Miyaji:2015fia, Nakayama:2015mva}, similar conditions were obtained 
for the special case of a bulk operator in the center of AdS by symmetry considerations.

In this section we will solve (\ref{cormodeq}) in coordinate space to recover the smearing function for a local bulk operator in the
complex coordinate representation.  We will do the same thing in momentum space and recover the bulk Poincar\'e modes
which make up a local bulk operator.  In addition we will solve the appropriate equations for a CFT at finite temperature and recover a local bulk operator in the BTZ background. This provides a new way of constructing the zeroth-order bulk operator and deriving bulk
modes without knowing anything about the bulk geometry.

Note however that the conditions for bulk locality (\ref{cormodeq}) only determine the bulk operator up to a coefficient.  The coefficient
could depend on bulk position, so in fact we can only generically recover
$\Phi^{(0)}$ up to a function of the bulk space-time coordinates. In states where the CFT has an unbroken spacetime translation symmetry the function can only depend on the bulk radial coordinate.  In this case dimensional analysis fixes $\Phi^{(0)}$ up
to an overall constant.  But in general locality is not enough to fix the function.  Even with this freedom we
get quite a lot of information. For example, given the two-point function of a local bulk operator with another local bulk or
boundary operator we can identify the singularities and deduce the bulk causal structure.\footnote{One can use this information
to reconstruct the bulk metric, up to a conformal factor, by the method of light-cone cuts \cite{Engelhardt:2016wgb}.} Also the program of
perturbatively correcting the zeroth-order bulk operator to take interactions into account only relies on the singularity structure, so up to a multiplicative coefficient an interacting local bulk operator could be constructed.  Moreover this multiplicative freedom
cancels in any ratio of correlation functions involving a fixed bulk operator with any number of  boundary operators, so one could determine
these ratios unambiguously. As another example of an unambiguous quantity, along the way we will see that the construction generates
the equations which describe bulk spacelike geodesics.

%%%%%%%%%%%%%%%%%%%%%%%%%
\subsection{Recovering smearing functions for the vacuum state}

We start with an ansatz for an object that commutes with the modular Hamiltonian
\begin{equation}
\Phi(X)=\int dt' dy' g(p,q){\cal O}(q,p)
\end{equation}
where $q=X-t'+iy'$, $p=X+t'+iy'$.  In the ansatz $t'$ and $y'$ are taken to be real and $X$ is left as a free real variable.
From (\ref{modhcomp}) the action of the modular Hamiltonian for a segment $(y_1,y_2)$ on $\Phi$ is given by
\begin{eqnarray}
& & [\tilde{H}_{mod}^{12},\Phi(X)]=\frac{2\pi i}{y_2-y_1} \int dt'dy' g(p,q) \times \\
& & \qquad \left ( (p-q)\Delta -y_1y_2(\partial_{q}-\partial_{p})+(y_1+y_2)(q \partial_{q}-p\partial_{p})+p^{2}\partial_{p}- q^2\partial_{q}\right ) {\cal O}(q,p)\nonumber
\label{modhcomp1}
\end{eqnarray}
We take two such modular Hamiltonians with parameters $(y_1,y_2)$ and $(y_3,y_4)$ and demand 
\begin{equation}
(y_2-y_1)[\tilde{H}_{mod}^{12},\Phi(\xi, \bar{\xi})]=0, \ \ (y_4-y_3)[\tilde{H}_{mod}^{34},\Phi(\xi, \bar{\xi})]=0
\label{basicintham}
\end{equation}
It's convenient to first look at the difference of the equations in (\ref{basicintham}),
\begin{equation}
 \int dt'dy' g(p,q)  ( (y_3y_4-y_1y_2)(\partial_{q}-\partial_{p})+(y_1+y_2-y_3-y_4)(q \partial_{q}-p\partial_{p})) {\cal O}(q,p)=0
\end{equation}
After integration by parts this gives an equation for $g(q,p)$,
\begin{equation}
((X_{0}-q)\partial_{q}-(X_{0}-p)\partial_{p})g(q,p)=0
\end{equation}
where $X_{0}=\frac{y_1y_2-y_3y_4}{y_1+y_2-y_3-y_4}$.  The solution to this equation is
\begin{equation}
g(q,p)=f((p-X_{0})(q-X_{0}))
\end{equation}
where $f$ is an arbitrary function.
We now use this form and solve the equation $[\tilde{H}_{mod}^{12},\Phi(X)]=0$. Following the same steps as before we get an equation for $f$
\begin{equation}
((\Delta-2)(p-q)+y_1y_2(\partial_{q}-\partial_{p})-(y_1+y_2)(q\partial_{q}-p\partial_{p})+q^2\partial_{q}-p^2 \partial_{p})f((q-X_{0})(p-X_{0}))=0
\end{equation}
whose solution is 
\begin{equation}
f=c_{\Delta}\left(Z^2+(p-X_{0})(q-X_{0})\right)^{\Delta-2}
\end{equation}
where
\be
Z^2=(y_1+y_2)X_{0}-y_1y_2-X_{0}^{2}
\label{fs}
\ee
The two parameters appearing in the solution $X_{0}$, $Z$ can be identified as the coordinates of the local operator in the bulk.
Note, for example, that as $y_1,y_3 \rightarrow y_2$ we have $Z \rightarrow 0$ and $X_0 \rightarrow y_2$.  Comparing (\ref{fs})
to (\ref{poingeo}), note that we have recovered from the CFT the equation which describes a spacelike geodesic in the bulk.

For the integration by parts to work without any boundary terms we need the integration region to be bounded by
$Z^2+(p-X_{0})(q-X_{0})=0$. For this to be possible for real $(t',y')$ we see that we must have $X=X_{0}$. So finally we get
\begin{equation}
\Phi(Z,X_{0})=c_{\Delta}\int_{t'^2+y'^2 < Z^2} dt' dy' (Z^2-t'^2-y'^2)^{\Delta-2}{\cal O}(t',X_{0}+iy').
\end{equation}
Since the vacuum state is translation invariant we expect correlation functions of local bulk fields to be translation invariant as well.
From this we can deduce that the coefficient $c_{\Delta}$ is a function of $Z$ only, which could be determined from a normalization condition
such as $\Phi({Z \rightarrow 0},X) \rightarrow \frac{Z^{\Delta}}{2\Delta-d}{\cal O}(X)$.
In this way we have recovered the bulk operator written in the complex coordinate representation.

%%%%%%%%%%%%%%%%%%%%%%%%%%%%
\subsubsection{Derivative representation}

Another possible representation for a bulk operator is 
\begin{equation}
\Phi(Z,\xi,\bar{\xi})=\sum_{n,m=0}^{\infty} a_{nm} \partial^n_{\xi}\partial^m_{\bar{\xi}} {\cal O}(\xi,\bar{\xi})
\label{phicor1}
\end{equation}
In this case formally we can impose locality using the usual modular Hamiltonian and we do not need the extended modular Hamiltonian.

We wish to impose the conditions (\ref{basicintham}), that $\Phi$ commutes with two modular Hamiltonians. As before we start by
looking at the difference of the two equations in (\ref{basicintham}) which gives
\begin{equation}
\sum_{n,m=0}^{\infty} a_{nm}\partial^n_{\xi}\partial^m_{\bar{\xi}}((\xi-X_{0})\partial_{\xi}+(X_{0}-\bar{\xi})\partial_{\bar{\xi}}){\cal O}(\xi,\bar{\xi})
\label{corderiv}
\end{equation}
Without loss of generality we take
\begin{equation}
\Phi(X_0)=\sum_{n,m=0}^{\infty} a_{nm}\partial^n_{\xi}\partial^m_{\bar{\xi}} {\cal O}(\xi,\bar{\xi})_{\xi=\bar{\xi}=X_{0}}
\end{equation}
Using this in (\ref{corderiv}) and setting the coefficients of $\partial^n_{\xi}\partial^m_{\bar{\xi}} {\cal O}(\xi=X_{0},\bar{\xi}=X_{0})$ to zero
gives
\begin{equation}
(n-m)a_{nm}=0
\end{equation}
So in fact $\Phi$ must have the form
\begin{equation}
\Phi(X_{0})=\sum_{m=0}^{\infty} a_{m}(\partial_{\xi}\partial_{\bar{\xi}})^m  {\cal O}(\xi,\bar{\xi})_{\xi=\bar{\xi}=X_{0}}
\label{phicor}
\end{equation}
Now demanding that
\begin{equation}
[\tilde{H}_{mod}^{12},\Phi(X_{0})]=0
\label{comeq}
\end{equation}
gives the condition
\begin{equation}
\sum_{n}a_m (m(\Delta+m-1)\partial^m_{\xi}\partial^{m-1}_{\bar{\xi}}-m(\Delta+m-1)\partial^{m-1}_{\xi}\partial^{m}_{\bar{\xi}}+Z^2\partial^{m+1}_{\xi}\partial^{m}_{\bar{\xi}}-Z^2\partial^{m}_{\xi}\partial^{m+1}_{\bar{\xi}}) {\cal O}(\xi,\bar{\xi})_{\xi=\bar{\xi}=X_{0}}=0
\label{finder}
\end{equation}
where $Z^2=-y_1y_2+(y_1+y_2)X_{0}-X_{0}^2$.  This implies the recursion relation 
\begin{equation}
a_{m}=-\frac{Z^2}{m(\Delta+m-1)}a_{m-1}
\end{equation}
whose solution is
\begin{equation}
\label{am}
a_{m}=a_{0}\frac{(-1)^m Z^{2m}}{\Gamma(m+1)\Gamma(\Delta +m)}
\end{equation}
As before time and space translation invariance restrict $a_{0}$ to be a function of $Z$.
The expression for a bulk operator in Poincar\'e coordinates (\ref{comprepcor}) can be expanded in derivatives \cite{daCunha:2016crm}.
\be
\Phi(Z,X,T) = \Gamma(\Delta)Z^{\Delta}\sum_{m=0}^{\infty} \frac{(-1)^m Z^{2m}}{\Gamma(m+1)\Gamma(\nu +m+1)}(\partial_{\xi} \partial_{\bar{\xi}})^m{\cal O}(\xi,\bar{\xi})
\ee
Comparing this to (\ref{phicor}) and (\ref{am}) we see that we have recovered the local bulk operator $\Phi(Z,X,T=0)$.

%%%%%%%%%%%%%%%%%%%%%%%%%
\subsection{Recovering bulk modes}

In this section we wish to recover the momentum space representation for a bulk operator, i.e.\ the bulk modes.
We start with the extended modular Hamiltonian for the segment $(y_1,y_2)$
\begin{equation}
[ \tilde{H}_{mod}, {\cal O}(\xi ,\bar{\xi})]=\frac{2\pi i}{y_2-y_1}\left ( (\bar{\xi}-\xi)\Delta -y_1y_2(\partial_{\xi}-\partial_{\bar{\xi}})+(y_1+y_2)(\xi \partial_{\xi}-\bar{\xi}\partial_{\bar{\xi}})+\bar{\xi}^{2}\partial_{\bar{\xi}}-\xi ^2\partial_{\xi}\right ) {\cal O}
\label{modpointot12}
\end{equation}
We define $k_{+}=\frac{k+\omega}{2}$, $k_{-}=\frac{\omega-k}{2}$ and 
\begin{equation}
{\cal O}(k_{+},k_{-})=\frac{1}{4\pi^2}\int d\xi d\bar{\xi} e^{-ik_{+}\xi+ik_{-}\bar{\xi}}{\cal O}(\xi ,\bar{\xi})
\end{equation}
Using (\ref{modpointot12}) one finds
\begin{eqnarray}
[\tilde{H}_{mod}, {\cal O}(k_{+},k_{-})]&=& \frac{2\pi i}{y_2-y_1}\left ( (y_1+y_2)(\frac{d}{dk_{-}}k_{-}-\frac{d}{dk_{+}}k_{+})
-2ih(\frac{d}{dk_{+}}+ \frac{d}{dk_{-}}) \right. \nonumber \\
&+& \left. \frac{d^2}{d^2 k_{+}} ik_{+}-\frac{d^2}{d^2 k_{-}} ik_{-} -y_1y_2(ik_{+}+ik_{-}) \right ){\cal O}(k_{+},k_{-})
\label{totmodmom}
\end{eqnarray}

We now look for operators $\Phi$ which commute with the extended modular Hamiltonians for two segments
$(y_1,y_2)$ and $(y_3,y_4)$.
\begin{equation}
(y_2-y_1)[\tilde{H}_{mod}^{12}, \Phi]=0, \ \  (y_4-y_3)[\tilde{H}_{mod}^{34}, \Phi]=0
\label{Hteq}
\end{equation}
We make the ansatz
\begin{equation}
\Phi=\int dk_{+}dk_{-} \, g(k_{+},k_{-}) {\cal O}(k_{+},k_{-})
\end{equation}
We first require that $\Phi$ satisfy the difference of the two equations in (\ref{Hteq}). This gives
\begin{equation}
\hspace{-9mm}\int dk_{+}dk_{-} \, g(k_{+},k_{-})\left((y_1+y_2-y_3-y_4)(\frac{d}{dk_{-}}k_{-}-\frac{d}{dk_{+}}k_{+})
-(y_1y_2-y_3y_4)(ik_{+}+ik_{-})\right) {\cal O}(k_{+},k_{-})=0
\end{equation}
Upon integration by parts we get an equation for $g(k_{+},k_{-})$,
\begin{equation}
(k_{+}\frac{d}{dk_{+}}-k_{-}\frac{d}{dk_{-}})g(k_{+},k_{-})=iX_{0}(k_{+}+k_{-})g(k_{+},k_{-})
\end{equation}
where $X_{0}=\frac{y_1y_2-y_3y_4}{y_1+y_2-y_3-y_4}$. The general solution to this equation is 
\begin{equation}
g(k_{+},k_{-})=f(k_{+}k_{-})e^{i(k_{+}-k_{-})X_{0}}
\end{equation}
where $f$ is an arbitrary function of $k_{+}k_{-}$.

Having imposed that $\Phi$ commutes with the difference $(y_2-y_1)\tilde{H}_{mod}^{12} - (y_4-y_3)\tilde{H}_{mod}^{34}$, we now require that $\Phi$ commute with $\tilde{H}_{mod}^{12}$ itself.  We start with the ansatz
\begin{equation}
\Phi(X_{0})=\int dk_{+}dk_{-}f(k_{+}k_{-})e^{i(k_{+}-k_{-})X_{0}}{\cal O}(k_{+},k_{-})
\label{bulkopmom}
\end{equation}
The first equation in (\ref{Hteq}) becomes a condition on $f$. After integration by parts and some algebra
we find ($x=k_{+}k_{-}=\frac{1}{4}(\omega^2-k^2)$)
\begin{equation}
x\frac{d^2 f}{d^2 x}+2h\frac{df}{dx}+((y_1+y_2)X_{0}-y_1y_2-X_{0}^{2})f=0
\label{feqn}
\end{equation}
The solution is 
\begin{equation}
f(\omega,k)=c_{0}(Z^{2}(\omega^2-k^2))^{-\nu/2}J_{\nu}(Z\sqrt{\omega^2-k^2})
\end{equation}
with $\Delta=\nu+1$ and $Z^2=(y_1+y_2)X_{0}-y_1y_2-X_{0}^{2}$.  Time and space translation invariance restrict $c_{0}$ to be a
function of $Z$.
Then (\ref{bulkopmom}) becomes\footnote{We could choose the coefficient $c_{0}$ so that $\Phi$ has the right limit as $Z \rightarrow 0$.
Also note that the boundary operator only has modes with $\vert \omega \vert > \vert k \vert.$}
\begin{equation}
\Phi(X_{0})=c_{0}\int d\omega dk \, e^{ikX_{0}}\big(Z^2(\omega^2-k^2)\big)^{-\nu/2}J_{\nu}(Z\sqrt{\omega^2-k^2}){\cal O}(k_{+},k_{-})
\end{equation}
which is the bulk operator $\Phi(Z,X_{0},T=0)$ given in (\ref{pomom}).

If we had imposed the second equation in (\ref{Hteq}) we would have gotten the same result with
$Z^2=(X_{0}-y_3)(y_4-X_{0})$.  But in fact these two expressions for $Z$ are the same. As long as the parameters $(y_1,y_2,y_3,y_4)$ assign a real value to $Z$  we have a solution where the point $(T=0,Z,X_{0})$ is the bulk point located at the intersection of the two boundary-anchored geodesics.

%%%%%%%%%%%%%%%%%%%%%%%%%%%%%%%%%%%%%%%%%%%%
\subsection{Time dependence}

For completeness we show that the construction of local bulk operators based on intersecting modular Hamiltonians
also captures the correct time dependence of bulk fields.

To do this we look at the modular Hamiltonian for a diamond that is shifted in time by an amount $T$.
In light-front coordinates (\ref{uv}) such a diamond is characterized by
\begin{equation}
u=y_2-T,\ \bar{u}=y_1+T, \  v=y_1-T, \  \bar{v}=y_2+T
\end{equation}
The extended modular Hamiltonian acting in momentum space is given by
\begin{eqnarray}
&&[\tilde{H}^{L}_{mod}(T), {\cal O}(k_{+},k_{-})]=\frac{1}{y_2-y_1}\left ((y_1+y_2-2T)(h-\frac{d}{dk_{+}}k_{+})\right. \\
&& \qquad - \left. 2ih\frac{d}{dk_{+}}+(\frac{d^2}{d^2 k_{+}} -(y_1 y_2-T(y_1+y_2)+T^2))ik_{+} \right ){\cal O}(k_{+},k_{-})\nonumber\\[5pt]
&&[\tilde{H}^{R}_{mod}(T), {\cal O}(k_{+},k_{-})]=\frac{1}{y_2-y_1}\left (-(y_1+y_2+2T)(h-\frac{d}{dk_{-}}k_{-})\right. \\
&& \qquad - \left. 2ih\frac{d}{dk_{-}}+(\frac{d^2}{d^2 k_{-}} -(y_1 y_2+T(y_1+y_2)+T^2))ik_{-} \right ){\cal O}(k_{+},k_{-})\nonumber
\end{eqnarray}
We look for operators that commute with $\tilde{H}^{total}_{mod}= \tilde{H}^{L}_{mod}+\tilde{H}^{R}_{mod}$.
We make an ansatz
\begin{equation}
\Phi(X)=\int dk_{+}dk_{-} \, g(k_{+},k_{-}){\cal O}(k_{+},k_{-})
\end{equation}
If we take two different boundary segments $(y_1,y_2)$ and $(y_3,y_4)$ at time $T$  then the conditions we wish to
impose are
\begin{equation}
(y_2-y_1)[\tilde{H}_{mod}^{12},\Phi]=0, \ \ (y_4-y_3) [\tilde{H}_{mod}^{34},\Phi]=0
\end{equation}
Taking the difference results in an equation for $g(k_{+},k_{-})$,
\begin{equation}
\left(k_{+}\frac{d}{dk_{+}}-k_{-}\frac{d}{dk_{-}}\right)g(k_{+},k_{-})=\big((X_{0}-T)ik_{+}+(X_{0}+T)ik_{-}\big)g(k_{+},k_{-})
\end{equation}
with $X_{0}=\frac{y_1y_2-y_3y_4}{y_1+y+2-y_3-y_4}$.  The solution to this equation is
\begin{equation}
g(k_{+},k_{-})=f(k_{+}k_{-})e^{i(k_{+}-k_{-})X}e^{-i(k_{+}+k_{-})T}
\end{equation}
with $f$ an arbitrary function of $k_{+}k_{-}$.
Thus our ansatz is now
\begin{equation}
\Phi(X_{0},T)=\int dk_{+}dk_{-} \, f(k_{+}k_{-})e^{i(k_{+}-k_{-})X}e^{-i(k_{+}+k_{-})T}{\cal O}(k_{+},k_{-})
\end{equation}
Having imposed the difference, the remaining condition
\begin{equation}
[\tilde{H}_{mod}^{12}(T), \Phi(X_{0},T)]=0
\end{equation}
is solved (after some algebra) by
\begin{equation}
f(k_{+}k_{-})\sim (Z^2(\omega^2-k^2))^{-\frac{\Delta-1}{2}}J_{\Delta-1}(Z\sqrt{\omega^2-k^2})
\end{equation}
where
\be
Z^2=(y_1+y_2)X_{0}-y_1 y_2 -X_{0}^2
\ee
Thus we've recovered the full Poincar\'e bulk mode, including its time dependence, purely from CFT considerations.
This shows that we can get the complete zeroth-order expression for a local bulk field using intersecting modular Hamiltonians.

%%%%%%%%%%%%%%%%%%%%%%%%%%%%
\subsection{Recovering BTZ bulk operators}

In this section we follow the same procedure to construct local bulk scalar fields in a BTZ background. We start with the extended modular Hamiltonian appropriate to a $(1+1)$-dimensional CFT at finite temperature.  The extended modular Hamiltonian for a segment has two parameters, the size of the segment $2L$ and the position of the center of the
segment $\phi_{0}$.
\begin{equation}
\hspace{-3mm}\tilde{H}_{mod,L,\phi_{0}}= c\int_{-\infty}^{\infty} \Big(\cosh \frac{r_{+} L}{l^2} -\cosh\frac{r_{+} (\xi-\phi_{0})}{l^2}\Big)T_{\xi \xi}(\xi)+c\int_{-\infty}^{\infty} \Big(\cosh \frac{r_{+} L}{l^2} -\cosh\frac{r_{+}(\bar{\xi}-\phi_{0})}{l^2}\Big)T_{\bar{\xi} \bar{\xi}}(\bar{\xi})
\label{hmodtherBTZ}
\end{equation}
We will consider two extended Hamiltonians
\be
\tilde{H}_{mod,L,\phi_{0}} \quad {\rm and} \quad \tilde{H}_{mod,R,\phi_{0}=0}
\ee
and wish to find a CFT operator $\Phi$ that satisfies
\be
[\tilde{H}_{mod,L,\phi_{0}},\Phi]=0, \ \ \ [\tilde{H}_{mod,R,\phi_{0}=0},\Phi]=0
\ee

The action of the extended Hamiltonian on a primary scalar operator of dimension $2h$ is
\begin{eqnarray}
[\tilde{H}_{mod,L,\phi_{0}},{\cal O}]&=&c\Big(-\frac{r_{+}h}{l^2}\sinh \frac{r_{+}(\xi-\phi_{0})}{l^2}+\frac{r_{+}h}{l^2}\sinh \frac{r_{+}(\bar{\xi}-\phi_{0})}{l^2} \\
&+&(\cosh \frac{r_{+}L}{l^2} -\cosh\frac{r_{+}(\xi-\phi_{0})}{l^2})\partial_{\xi}-(\cosh \frac{r_{+}L}{l^2} -\cosh\frac{r_{+}(\bar{\xi}-\phi_{0})}{l^2})\partial_{\bar{\xi}}\Big){\cal O}(\xi,\bar{\xi}) \nonumber
\end{eqnarray}
We define the variables as before 
\begin{equation}
q=l\phi-\frac{l^2 x}{r_{+}}+i\frac{l^2y}{r_{+}}, \qquad p=l\phi+\frac{l^2 x}{r_{+}}+i\frac{l^2y}{r_{+}}
\label{basicpq}
\end{equation}
where $\phi$ is a free parameter, and we define rescaled variables
\begin{equation}
\tilde{q}=\frac{r_{+}}{l^2}q, \ \ \ \tilde{p}=\frac{r_{+}}{l^2}p, \ \ \ \tilde{L}=\frac{r_{+}}{l^2}L, \ \ \ \tilde{\phi_{0}}=\frac{r_{+}}{l}\phi_{0}
\end{equation}
Starting with the general ansatz
\begin{equation}
\Phi=\int d\tilde{p}d\tilde{q} \, g(\tilde{p},\tilde{q}){\cal O}(\tilde{p},\tilde{q})
\end{equation}
the condition
\begin{equation}
[\tilde{H}_{mod,L,\phi_{0}},\Phi]=0
\end{equation}
becomes upon integration by parts\footnote{We will choose the region of integration to ensure that there are no boundary terms.}
\begin{eqnarray}
&& \left ((\cosh \tilde{L} - \cosh(\tilde{q}-\tilde{\phi_{0}}))\partial_{\tilde{q}}-(\cosh \tilde{L}-\cosh (\tilde{p}-\tilde{\phi_{0}}))\partial_{\tilde{p}}\right )g(\tilde{p},\tilde{q}) \nonumber\\
&& \qquad = (h-1)\big(\sinh (\tilde{p}-\tilde{\phi_{0}})-\sinh (\tilde{q}-\tilde{\phi_{0}})\big)g(\tilde{p},\tilde{q})
\label{genhamrin}
\end{eqnarray}

We first impose the condition $[\tilde{H}_{mod,R,\phi_{0}=0},\Phi]=0$.  To do this we set
$\tilde{\phi_{0}}=0$ and $\tilde{L}=\tilde{R}$ in (\ref{genhamrin}).  Then
using the method of characteristics, the most general solution to (\ref{genhamrin}) is
\begin{eqnarray}
&& g(\tilde{p},\tilde{q}) = c_{0}f(x)K^{h-1}\nonumber\\[3pt]
&& x = \frac{\sinh (\frac{\tilde{R}+\tilde{q}}{2})\sinh (\frac{\tilde{R}+\tilde{p}}{2})}{\sinh (\frac{\tilde{R}-\tilde{q}}{2})\sinh (\frac{\tilde{R}-\tilde{p}}{2})} \nonumber\\[3pt]
&& K = \sinh (\frac{\tilde{R}+\tilde{q}}{2})\sinh (\frac{\tilde{R}+\tilde{p}}{2})\sinh (\frac{\tilde{R}-\tilde{q}}{2})\sinh (\frac{\tilde{R}-\tilde{p}}{2})
\label{ansrin}
\end{eqnarray}
where $f$ is an arbitrary function and $c_{0}$ is a constant.
Since we also want $\Phi$ to obey
\begin{equation}
[\tilde{H}_{mod,L,\phi_{0}},\Phi]=0
\end{equation}
we re-insert the solution (\ref{ansrin}) into (\ref{genhamrin}).  This now gives an equation for $f(x)$.
After some algebra the equation can be recast as
\begin{eqnarray}
&& \frac{df}{dx} = \frac{h-1}{x}\frac{x-\alpha}{x+\alpha}f \nonumber\\[3pt]
&& \alpha = \frac{\sinh \tilde{\phi_{0}} \sinh \tilde{R}+\cosh \tilde{\phi_{0}} \cosh \tilde{R}-\cosh \tilde{L}}{\sinh \tilde{\phi_{0}} \sinh \tilde{R}-\cosh \tilde{\phi_{0}} \cosh \tilde{R}+\cosh \tilde{L}}
\end{eqnarray}
with solution
\begin{equation}
f(x)=c_1 \left (\frac{(x+\alpha)^2}{x} \right )^{h-1}
\end{equation}
The parameter $\alpha$ can be seen to depend on only two parameters by defining
\begin{equation}
\tanh \tilde{\phi_{*}}=\frac{1}{\sinh \tilde{\phi_{0}}}\Big(\cosh \tilde{\phi_{0}}-\frac{\cosh \tilde{L}}{\cosh \tilde{R}}\Big)
\label{rininter}
\end{equation}
and noting that
\begin{equation}
\alpha=\frac{\cosh \tilde{\phi_{*}} \sinh \tilde{R}+\sinh \tilde{\phi_{*}} \cosh \tilde{R}}{\cosh \tilde{\phi_{*}} \sinh \tilde{R}-\sinh \tilde{\phi_{*}} \cosh \tilde{R}}
\end{equation}
We can set the free parameter $\phi$ in (\ref{basicpq}) to be $\phi=\phi_{*}$ so that
$g(\tilde{p},\tilde{q})$ becomes
\begin{equation}
g(\tilde{p},\tilde{q})=c_2\left( \cosh (\frac{\tilde{p}+\tilde{q}}{2}-\tilde{\phi_{*}})-\frac{\cosh\tilde{\phi_{*}}}{\cosh \tilde{R}}\cosh (\frac{\tilde{p}-\tilde{q}}{2})\right )^{\Delta-2}=c_2 \left (\cos y - \frac{\cosh\tilde{\phi_{*}}}{\cosh \tilde{R}} \cosh x \right )^{\Delta-2}
\label{lastrinsm}
\end{equation}
The two parameters of the solution $\tilde{\phi_{*}}$ and $\frac{\cosh\tilde{\phi_{*}}}{\cosh \tilde{R}}$ can be identified as the coordinate parallel to the boundary and the radial coordinate, respectively, by looking at the limit $\tilde{L}\rightarrow 0$, $\tilde{\phi_{0}}\rightarrow \tilde{R}$. In this limit 
$\tilde{\phi_{*}} \rightarrow \tilde{R}$ and $\frac{\cosh\tilde{\phi_{*}}}{\cosh \tilde{R}}\rightarrow 1$ as expected.

The region of integration is fixed by requiring that there are no boundary terms when we integrate by parts.  This determines
the region of integration to be
\begin{equation}
\cos y > \frac{\cosh\tilde{\phi_{*}}}{\cosh \tilde{R}} \cosh x
\end{equation}
On an equal-time geodesic stretching from $-\tilde{R}$ to $\tilde{R}$, we show in appendix C that the boundary coordinate $\phi$
and the bulk coordinate $r$ are related by
\begin{equation}
\sqrt{1-\frac{r^{2}_{+}}{r^2}}=\frac{\cosh \tilde{\phi}}{\cosh \tilde{R}}
\end{equation}
Thus (\ref{lastrinsm}) is the smearing function for a bulk scalar operator $\Phi(r,\phi_{*}, t=0)$ in a BTZ background \cite{Hamilton:2006az}.
In appendix C we show that $\phi_*$ in (\ref{rininter}) is just the $\phi$ coordinate where the bulk geodesics intersect.
Thus again we have recovered the bulk space-like geodesics from the CFT.

%%%%%%%%%%%%%%%%%%%%%%%%%
\section{Conclusions}

In this paper we have shown that CFT operators which mimic local bulk operators commute with the modular Hamiltonian appropriate for a boundary segment whose RT surface passes through the bulk point.  If two RT surfaces intersect at a point in the bulk then a bulk observable localized on the intersection must commute with both modular Hamiltonians.  Turning this around, we used this as a new
way to construct local bulk observables in the CFT, by constructing CFT quantities which commute with intersecting extended modular Hamiltonians. Along the way we recovered bulk space-like geodesics from the CFT.

The computations done in this paper were for AdS${}_{3}$ / CFT${}_2$, but the generalization to higher dimensions is clear.  The only complication is that AdS${}_D$ requires $D-1$ intersecting
RT surfaces to define a bulk point.

It seems clear that at least in principle the construction can be carried out for CFT states which are not the vacuum. Indeed in this paper the finite temperature case was treated successfully.  Explicit expressions may be difficult to obtain since we have little control
over the modular Hamiltonian for non-vacuum states.  Moreover in general the modular Hamiltonian will be non-local, so (unlike the examples treated in this paper) for generic states the approach will not lead to a system of local differential equations for the smearing functions.  But in principle the same logic applies and should determine local bulk operators in the appropriately-deformed bulk background
geometry.

Another, perhaps related, generalization of the construction in this paper would be to include interactions and make contact with the perturbative
procedure developed in \cite{Kabat:2011rz, Kabat:2015swa,Kabat:2016zzr}.  It would also be interesting to understand if there is a connection to the ideas proposed in \cite{Lewkowycz:2016ukf, Guica:2016pid}.

In this paper we only considered scalar operators. It would be interesting to extend the construction to bulk fields with spin.  For massive vector fields this seems straightforward.  Bulk fields with gauge redundancy pose an additional challenge, since due to
constraints they aren't local objects in the bulk even at the free field level.\footnote{For a recent treatment of observables
for gauge fields see \cite{Donnelly:2015hta}.} Moreover even for bulk scalars gravitational dressing arises as an interaction effect, and once this is taken into account one cannot localize bulk scalar observables to the intersection of RT surfaces.  This means
that for free bulk gauge fields, and for interacting bulk scalars, one cannot simply demand that bulk observables commute with intersecting modular Hamiltonians.  Whether there is an extension of the approach to deal with
these issues is an interesting and important question.

The construction developed in this paper raises more speculative issues as well.  For example it seems clear that the construction puts constraints on CFT states which are dual to classical bulk geometries.
This comes about because a classical bulk geometry requires that an infinite family of equations, stating that different modular Hamiltonians
commute with a smeared CFT operator, must all have a common solution. This restricts the form of the modular Hamiltonians and hence presumably
the CFT states that can be dual to classical geometries. It would be interesting to make these restrictions more precise.

\bigskip
\goodbreak
\centerline{\bf Acknowledgements}
\noindent
We are grateful to Michal Heller for valuable comments.
The work of DK is supported by U.S.\ National Science Foundation grant PHY-1519705.  The work of GL is supported in part by the Israel Science
Foundation under grant  504/13.

%%%%%%%%%%%%%%%%%%%%%%%%%%%%%%%%%%%%%%%%%%
\appendix
\section{Extended modular Hamiltonian}
Here we show that given an interval $A$ and its complement $\bar{A}$, the extended modular Hamiltonian for the region $A$ can be identified as
\begin{equation}
\tilde{H}_{mod,A}=H_{mod, A}-H_{mod,\bar{A}}
\label{extmodn}
\end{equation}
This generalizes the usual extension of a Rindler time translation outside the Rindler wedge and is analogous to defining a thermofield Hamiltonian.

It's straightforward to show (\ref{extmodn}) for the vacuum state of a CFT on a line.
In \cite{Cardy:2016fqc} the modular Hamiltonian for a region was constructed using a conformal transformation with an analogy to electrostatics. Given the appropriate conformal transformation $f(z)$ the modular Hamiltonian for a region $A$ was given by
\begin{equation}
H_{mod}^{(R)}=2\pi \int_{A}\frac{T_{zz}(z)}{f'(z)} dz
\end{equation}
and similar for $H_{mod}^{(L)}$.  If we take the region $B$ to be the union of the segments $(-\infty, y_1)$ and $(y_2,\infty)$, then the appropriate $f(z)$ is just $f(z)=\ln \left(\frac{y_2-z}{z-y_1}\right )$ and similar for $\bar{f}(\bar{z})$.\footnote{For a general region consisting of two segments this is not a correct procedure since $f'(z)$ vanishes somewhere in the complex plane.  See the discussion section in \cite{Cardy:2016fqc}. However for the two semi-infinite segments we are considering this problem does not arise.} This gives the right-moving part of the modular Hamiltonian for this region to be 
\begin{equation}
H_{mod,B}^{(R)}=-2\pi \int_{-\infty}^{y_1} \frac{(y_2-z)(z-y_1)}{y_2-y_1}T_{zz}(z)-2\pi\int_{y_2}^{\infty} \frac{(y_2-z)(z-y_1)}{y_2-y_1}T_{zz}(z).
\end{equation}
Together with (\ref{Hmoddef}) and (\ref{Extmoddef}) this establishes (\ref{extmodn}) for the vacuum state.

%%%%%%%%%%%%%%%%%%%%%%%%%%%%%%%%%%%%%%%%%%
\section{Action of $\tilde{H}_{mod}$ on operators off the RT surface}
We make the ansatz
\begin{equation}
\Phi_{12}(X,T=0)=\int dk_{+}dk_{-} \, f(k_{+}k_{-})e^{i(k_{+}-k_{-})X}{\cal O}(k_{+},k_{-})
\end{equation}
where $f$ solves
\begin{equation}
k_{+}k_{-}\frac{d^2 f}{d(k_{+}k_{-})^2}+2h\frac{df}{d k_{+}k_{-}}+((y_1+y_2)X-y_1 y_2 -X^2)f=0
\end{equation}
This is the condition (\ref{feqn}) that $\Phi_{12}$ commutes with the extended modular Hamiltonian of the segment $(y_1,y_2)$.
Now consider the extended modular Hamiltonian $\tilde{H}_{mod}^{34}$ for a different segment $(y_3,y_4)$.
We wish to compute the commutator of this new modular Hamiltonian with $\Phi_{12}$. A simple computation gives
\begin{eqnarray}
&& [\tilde{H}^{34}_{mod},\Phi_{12}(Z,X,T=0)]=\frac{2\pi i}{(y_4-y_3)}((y_3+y_4-y_1-y_2)X-y_3 y_4 +y_1 y_2) \times \nonumber \\
&& \hspace{3cm} \int dk_{+}dk_{-} \, f(k_{+}k_{-})i(k_{+}+k_{-})e^{i(k_{+}-k_{-})X}{\cal O}(k_{+},k_{-})
\end{eqnarray}
which is just
\begin{equation}
[\tilde{H}^{34}_{mod},\Phi_{12}(Z,X,T=0)]=\frac{2\pi i}{(y_4-y_3)}\left((y_3+y_4-y_1-y_2)X-y_3 y_4 +y_1 y_2\right)\partial_{T}\Phi(X,T=0)
\label{locdev}
\end{equation}
Defining $y_4-y_3=2Z_{0}$, $y_4+y_3=2X_{*}$  we get
\begin{equation}
[\tilde{H}^{34}_{mod},\Phi_{12}(Z,X,T=0)]=-\frac{i\pi }{Z_{0}}\left((Z_{0}^{2}-Z^2-(X-X_{*})^2\right)\partial_{T}\Phi(Z,X,T=0)
\end{equation}
which is the correct action of the bulk Rindler Hamiltonian associated with the segment $(X_{*}-Z_{0},Z_{0}+X_{*})$,
i.e.\ it generates a Rindler time translation.

%%%%%%%%%%%%%%%%%%%%%%%%%%%%%%%
\section{Geodesics in BTZ}

The BTZ metric is 
\begin{equation}
ds^2=-\frac{r^2-r^{2}_{+}}{l^2}dt^2+\frac{l^2}{r^2-r_{+}^2}dr^2 +r^2 d\phi^2
\end{equation}
We look for geodesics $r(\phi)$ which extremize the action
\begin{equation}
\int d\phi \, \sqrt{r^2+\frac{l^2}{r^2-r_{+}^2} \left(\frac{dr}{d\phi}\right)^2}
\end{equation}
Since nothing depends explicitly on $\phi$ there is a constant of motion which we call $r_{min}$.
\begin{equation}
r_{min}=\frac{r^2}{\sqrt{r^2+\frac{l^2}{r^2-r_{+}^2} (\frac{dr}{d\phi})^2}}
\end{equation}
If we choose $r(\phi_{0})=r_{min}$ and require $\phi(r \rightarrow  \infty)=\pm L/l$  the solution after a little algebra is
\begin{equation}
\frac{ \cosh \frac{r_{+}}{l}(\phi-\phi_{0})}{\cosh \frac{r_{+}}{l^2}L}=\sqrt{1-\frac{r_{+}^{2}}{r^2}}
\end{equation}
Thus two geodesics, one stretching from $-R$ to $R$ and the other from $\phi_{0}-L$ to $\phi_{0}+L$, intersect in the bulk at a point whose
$\phi$ coordinate obeys
\begin{equation}
\tanh \frac{r_{+}}{l}\phi =\frac{1}{\sinh \frac{r_{+}}{l}\phi_{0}}\Big(\cosh \frac{r_{+}}{l}\phi_{0}-\frac{\cosh \frac{r_{+}}{l^2} L}{\cosh \frac{r_{+}}{l^2}R}\Big)
\end{equation}

%\bibliographystyle{utphys}
%\bibliography{gravity}
%%%%%%%%%%%%%%%%%%%%%%%%%%%%%%%%%%%%%%%
\providecommand{\href}[2]{#2}\begingroup\raggedright\endgroup

\end{document}